\documentclass{article}

\usepackage{amsmath}
\usepackage{amsfonts}
\usepackage{amssymb}
\usepackage{graphicx}
\usepackage{hyperref}
\usepackage[utf8]{inputenc}
\usepackage{wrapfig}
\usepackage{caption}
\usepackage{eurosym}
\usepackage{lscape}
\usepackage{appendix}
\usepackage{wrapfig}
\usepackage[absolute]{textpos} 
\usepackage{fancyhdr}
\usepackage{geometry}
\date{}
\begin{document}

\title{Simulating coliform transport and decay from 3D hydrodynamics model and in situ observation in Nice area}


\author{Rémi Dumasdelage,\\ Metropole Nice Côte d'Azur and Nice Municipality, France \\
        Olivier Delestre, \\ Laboratoire J.A. Dieudonné UMR CNRS 7351 and Polytech Nice Sophia, Université Côte d'Azur
}


\maketitle

\begin{abstract}
In coastal area, raw wastewater can be directly discharged in the sea during extreme rain events or failure of the sewage network. The highly urbanized and touristic area of Nice is directly exposed to this risk.
In order to protect the bathing population and to apply the European regulation (DCE, 2000), Nice municipality has decided to work on a forecasting tool of transport and fate of fecal bacterial indicators such as Escherichia Coli.\\

This paper deals with the first step of this approach which consists in developing a three-dimensional   hydrodynamic simulation coupled with water quality parameters. The model used is based on Reynolds-Averaged Navier-Stokes's equations for the calculation of the velocities and water depth (TELEMAC3D), and wave energy balance equation to account for the wave induced current (TOMAWAC).
Two field measurements were conducted on a real wastewater pollution coming from a sewage pipe located at -38m depth. 

The first campaign has validated the vertical diffusion of the effluent through the water column by the means of variable depth samples and has stressed the hourly variability of sewage bacterial quality. Due to the lack of data regarding this last point, the model could not be fully validated. 
The second campaign has been designed to overcome this issue and to gather all the data needed to proceed to the validation of our hydro-ecological simulation. The comparison of the field measurements and the simulations has successfully demonstrated the accuracy of the model. But the uncertainty associated with the laboratory analysis (MPN) are high, making the calibration and validation processes a complex task.

Nevertheless, further research needs to be done regarding the variability of microbial quality in order to develop an operational tool allowing the forecast of bacterial pollution. The next step of this study will be to establish the hourly variation of microbial quality in the wastewater during dry and raining periods. \\

\end{abstract}

\section{Introduction}
\label{}

\begin{figure}
\centering
\includegraphics[width=14cm]{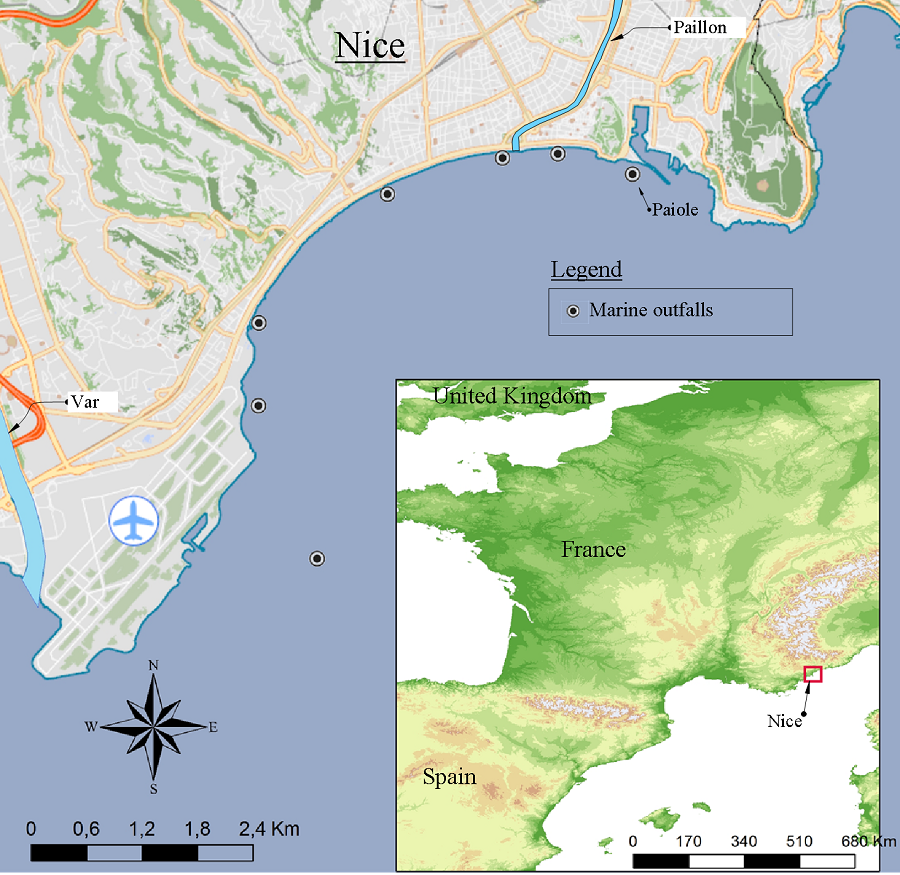}
\caption{Location of Nice and its 7 marine outfalls}
\label{situation}
\end{figure}

Nice is a highly urbanized town of the French Riviera (population: 500 000) with an economy mainly based on tourism (fig. \ref{situation}). Famous for its gravel beach and its sea front called Promenade des Anglais, this area hosts around 4.3 millions of tourists during the peak season in summer  with a number of people on the beach estimated to 20 000 persons per day during July and August.

Nice coast is composed of 4.5 km of pebble cordon divided  by a coastal river mouth called le Paillon located on figure \ref{situation}.  Under the West part of the beach and the East part of the sea front the main pipe of the town sewer system is buried 10 feet underground and transports sewage to a treatment plant.

The majority of Nice sewer system is handling both sanitary waste and storm water. During heavy rainfalls the sewer can be overloaded and the overflowing untreated water is discharged through seven marine outfalls. The sewage discharged in the sea can lead to bacterial pollution events. In this geographical area, such events were never monitored and simulated before the experiment presented in this paper.

The directive 2006/7/CE was enacted by the European Union to reduce the number of these events and so improve the bathing water quality \cite{EU06}. This legislative text lays down provisions for the monitoring, the classification, the communication and the management of bathing water quality. The directive was transposed to the French law in September 2008 by the Decree N2008-990 \cite{CSP08}. Two main points were developed for coastal waters:

$\bullet$ the first concerns the parameters monitored which are the microbiological indicators: Escherichia coli and intestinal enterococci are quantified by concentration values in colony forming unit per 100 ml (cfu/100ml);

$\bullet$ the second point is the bathing water which has to be classified as 'poor', 'sufficient', 'good', or 'excellent' based on the microbiological concentration  values (see tab. \ref{TabSeuil1} and \ref{TabSeuil2}). If  bathing water is classified as poor for five consecutive years, a permanent bathing prohibition or permanent advice against bathing shall be introduced.

\begin{table}[!h]
	\centering
		\begin{tabular}{|l|c|c|r|}
			\hline
			Intestinal Interococci & Excellent quality & Good quality & Sufficient \\
			\hline
			Legal thresholds  & 100 & 200 & 185 \\
			\hline
			Upper bound of the 95\% confidence interval & 217 & 351 & 330 \\
			\hline
			Lower bound of the 95\% confidence interval & 46 & 114 & 104 \\
			\hline			
		\end{tabular}
	\caption{Concentration thresholds for \textit{Intestinal Interococci} in cfu/100ml and the 95\% confidence interval from the most probable number of bacteria \cite{McGrady18}, \cite{Halvorson32}}
	\label{TabSeuil1}
\end{table}

\begin{table}[!h]
	\centering
		\begin{tabular}{|l|c|c|r|}
			\hline
			Escherichia Coli & Excellent quality & Good quality & Sufficient \\
			\hline
			Legal thresholds  & 250 & 500 & 500 \\
			\hline
			Upper bound of the 95\% confidence interval & 415 & 734 & 734 \\
			\hline
			Lower bound of the 95\% confidence interval & 150 & 340 & 340 \\
			\hline			
		\end{tabular}
	\caption{Concentration thresholds for \textit{Escherichia Coli} in cfu/100ml and the 95\% confidence interval from the most probable number of bacteria \cite{McGrady18}, \cite{Halvorson32}}
	\label{TabSeuil2}
\end{table}

In France, the municipalities are legally responsible for protecting  public health. As a consequence, the bathing water management falls directly within their jurisdiction \cite{CGCT}. The sewer system of Nice and its geographical situation make it sensible to the microbiological pollution events. During the last four years the city has faced 10 legal thresholds overun in the summer season. 
Thus, to preserve public health, to optimize the water bath quality, and   to stick with  the directive requirements,  Nice municipality has decided to develop its knowledge regarding the microbiological pollution.

The main objective of this paper is to present the first step of this project which consists in developing a three-dimensional hydrodynamic simulation of microbiological water quality parameters. Firstly,  the models used are introduced (i) as well as the field experiments carried out to validate them (ii). Then this paper reviews the results of these measurements (iii) and the accuracy of the 3D-simulations developed is discussed (iv).

\section{Method}

\subsection{\textbf{Reynolds-Averaged Navier-Stokes code coupled with a wave energy balance equations solver}}
The temperature and the salinity of the sea water can change rapidly through the water column \cite{Bedri11}.
In the Nice bay, these physical changes have been deeply studied by measurement campaigns \cite{Dolle03}. The variability of the temperature and the salinity are linked with meteorological parameters (wind, the exterior temperature) and hydrodynamic driving forces \cite{Muhammetoglu12}.
Thus, the use of 3D modelling tools was compulsary to represent the impact of this variabilty on the convection and diffusion transport of microbiological pollution \cite{Gourmelon10}.

TELEMAC-3D solves the Reynolds-Averaged Navier-Stokes equations  \cite{Hervouet07}, \cite{Pedlosky87}, \cite{Bedri11} with the incompressible form of the mass conservation equation (1) and the non-conservative momentum equations (2-4): 
\begin{align}
\frac{\partial{U}}{\partial x}+\frac{\partial{V}}{\partial{y}}+\frac{\partial{W}}{\partial z}&=0\\
\frac{\partial{U}}{\partial{t}}+\vec{U}.\vec{\nabla}(U)&=-\frac{1}{\rho_{0}}\frac{\partial{p}}{\partial{x}}+\frac{\partial}{\partial x}(\nu_{H}\frac{\partial{U}}{\partial x})+\frac{\partial}{\partial y}(\nu_{H}\frac{\partial{U}}{\partial y})+\frac{\partial}{\partial z}(\nu_{V}\frac{\partial{U}}{\partial z})+F_{x}\\
\frac{\partial{V}}{\partial{t}}+\vec{U}.\vec{\nabla}(V)&=-\frac{1}{\rho_{0}}\frac{\partial{p}}{\partial{y}}+\frac{\partial}{\partial x}(\nu_{H}\frac{\partial{V}}{\partial x})+\frac{\partial}{\partial y}(\nu_{H}\frac{\partial{V}}{\partial y})+\frac{\partial}{\partial z}(\nu_{V}\frac{\partial{V}}{\partial z})+F_{y}\\
\frac{\partial{W}}{\partial{t}}+\vec{U}.\vec{\nabla}(W)&=-g-\frac{1}{\rho_{0}}\frac{\partial{p}}{\partial{z}}+\frac{\partial}{\partial x}(\nu_{H}\frac{\partial{W}}{\partial x})+\frac{\partial}{\partial y}(\nu_{H}\frac{\partial{W}}{\partial y})+\frac{\partial}{\partial z}(\nu_{V}\frac{\partial{W}}{\partial z})+F_{z}
\end{align}
where $U$, $V$ and $W$ are the components of the velocity vector $\vec{U}$ in the $x$, $y$ and $z$ directions respectively; $p$ is the pressure at the depth $z$; $\rho_0$ is the reference density value; $\nu_{H}$ and $\nu_{V}$ are the horizontal and vertical eddy viscosity respectively; $g$ is the acceleration of gravity and $Fx$, $Fy$, $Fz$ are the external forces components such as Coriolis force, centrifugal force or the wind forcing, in the $x$, $y$ and $z$ directions respectively.
 
The non-hydrostatic pressure assumption is used in this study. Therefore, the acceleration and eddy viscosity terms for the vertical velocity component of the momentum equation are taken into account through the dynamic pressure $p_{d}$ \cite{Pedlosky87}. Hence, the total pressure $p$ can be expressed as the sum of this dynamic pressure $p_d$ which includes the vertical acceleration and the vertical viscosity term, and the hydrostatic term $p_h$ which represents the atmospheric pressure on the surface and the weight of the water column above it:

\begin{equation}
p = \underbrace{p_{atm}+\rho_{0}g(Z_{s}-z)+\rho_{0}g\int^{Z_{s}}_{z}\frac{\Delta\rho}{\rho_{0}}dz}_{p_h} + p_d,
\end{equation}

where $Z_s$ is the free surface elevation; $p_{atm}$ is the atmospheric pressure; $\rho$ is the density of water; 

The studied area is 60 $km^{2}$ wide and 200 $m$ deep, so the vertical length cannot be neglected compared to the horizontal scales \cite{Chen05}. This observation justifies the use of the non-hydrostatic assumption expressed previously.

The density varies with the sea water temperature $T$ and the salinity $S$. Even if these variations are considered as minor in the incompressible form of the mass conservation equation, these two parameters will be taken into account using the simplified equation of state established by UNESCO \cite{UNESCO87},\cite{Hervouet07}:

\begin{equation}
\rho=\rho_{ref}(1-(T(T-T_{ref})^2-750S)10^{-6}),
\end{equation}

where $T_{ref}$ is the reference temperature of $277.15K$ and $\rho_{ref}$ is the reference density at the temperature $T_{ref}$ when the salinity is zero.

The temperature and salinity are interacting with the hydrodynamics driving forces. Thus they are considered as active tracers $C_{r}$ and will be simulated through a non-conservative equation based on the mass conservation equation \cite{Hervouet07}:

\begin{equation}
\frac{\partial C_{r}}{\partial t}+(\vec{U}.\vec{\nabla})C_{r}=\vec{\nabla}.(\nu_{C_{r}} \vec{\nabla(C_{r})})+Q,
\end{equation}
 
where $Q$ is the source or sink term; $\nu_{T_{r}}$ is the diffusion coefficient of the tracer.
This equation is also used to model the mean concentration of Escherichia Coli  $C_{ec}$ (cfu/100ml).
\begin{equation}
\frac{\partial C_{ec}}{\partial t}+(\vec{U}.\vec{\nabla})C_{ec}=\vec{\nabla}.(\nu_{C_{ec}} \vec{\nabla(C_{ec})})+Q_{ec}.
\end{equation}

The decay of E.Coli is defined in the OPENTELEMAC by the sink term $Q_{ec}$ expressed as \cite{Bedri11}:
\begin{equation}
Q_{ec}=-\frac{2.303}{T_{90}}.C_{ec},
\end{equation}
where $T_{90}$ is the time expressed in second during which the bacterial population   reduces by  90\%.\\

Contrarily, the temperature and the salinity, E.Coli is a passive tracer. It has no interaction with the hydrodynamic of the flow.
The inactivation process of the fecal bacteria in the sea depends on  environmental  factors which are mainly the light intensity \cite{Feitosa13}, the temperature and the salinity \cite{Lopez13}, \cite{Muhammetoglu12}. Amongst the formulations established in this field \cite{Bellair77}, \cite{Chamberlin78}, \cite{Mancini78}, \cite{Guillaud97}, \cite{Yang00},  the formula proposed by Canteras \textit{et al.} \cite{Canteras95} was used to calculate the inactivation time:
\begin{equation}
T_{90}=\frac{2.303}{1.040^{T-20}1.012^{S}2.533+0.113I(\frac{c}{H})}
\label{equaCanter}
\end{equation}
where $c$ is the light extinction coefficient in the water, $H$ the water depth, and $I$ the light intensity

The experiment presented in this study was carried out during the night, so the light intensity terms has been neglected. The decay rate was calculated as 25 hours and supposed to be constant because of the little variability of the water column salinity and temperature presented in the next section (fig. \ref{Temp_salinity-2015}). In further applications of this model, a fortran subroutine has been implemented to change the inactivation time at each time step based on these environmental parameters.

\begin{figure}

	\centering
		\includegraphics[width=12cm]{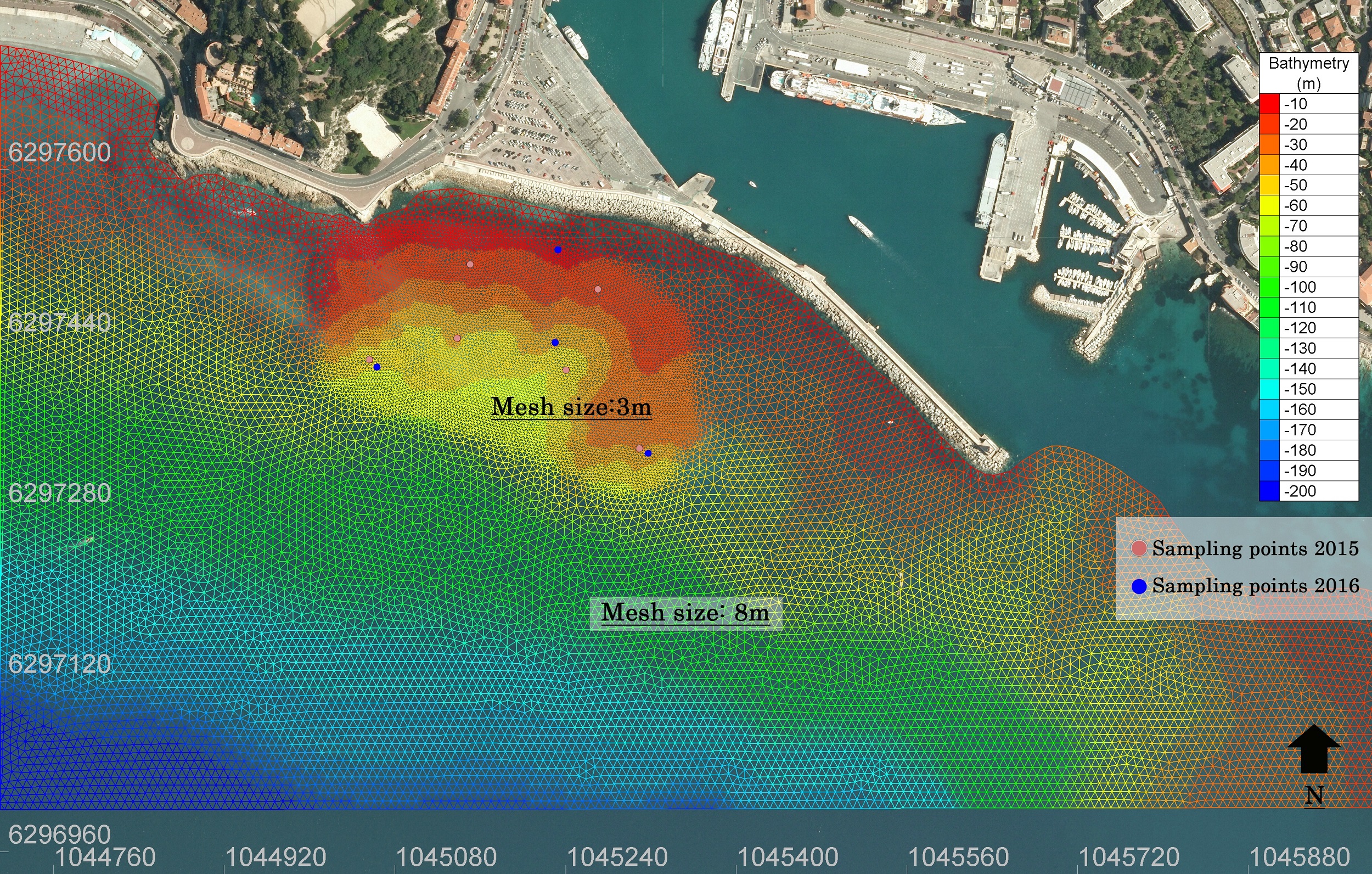}
	\caption{Computational 2D grid of Nice coastal area.}
	\label{Mesh2D}
\end{figure}

\begin{figure}
	\centering
		\includegraphics[width=12cm]{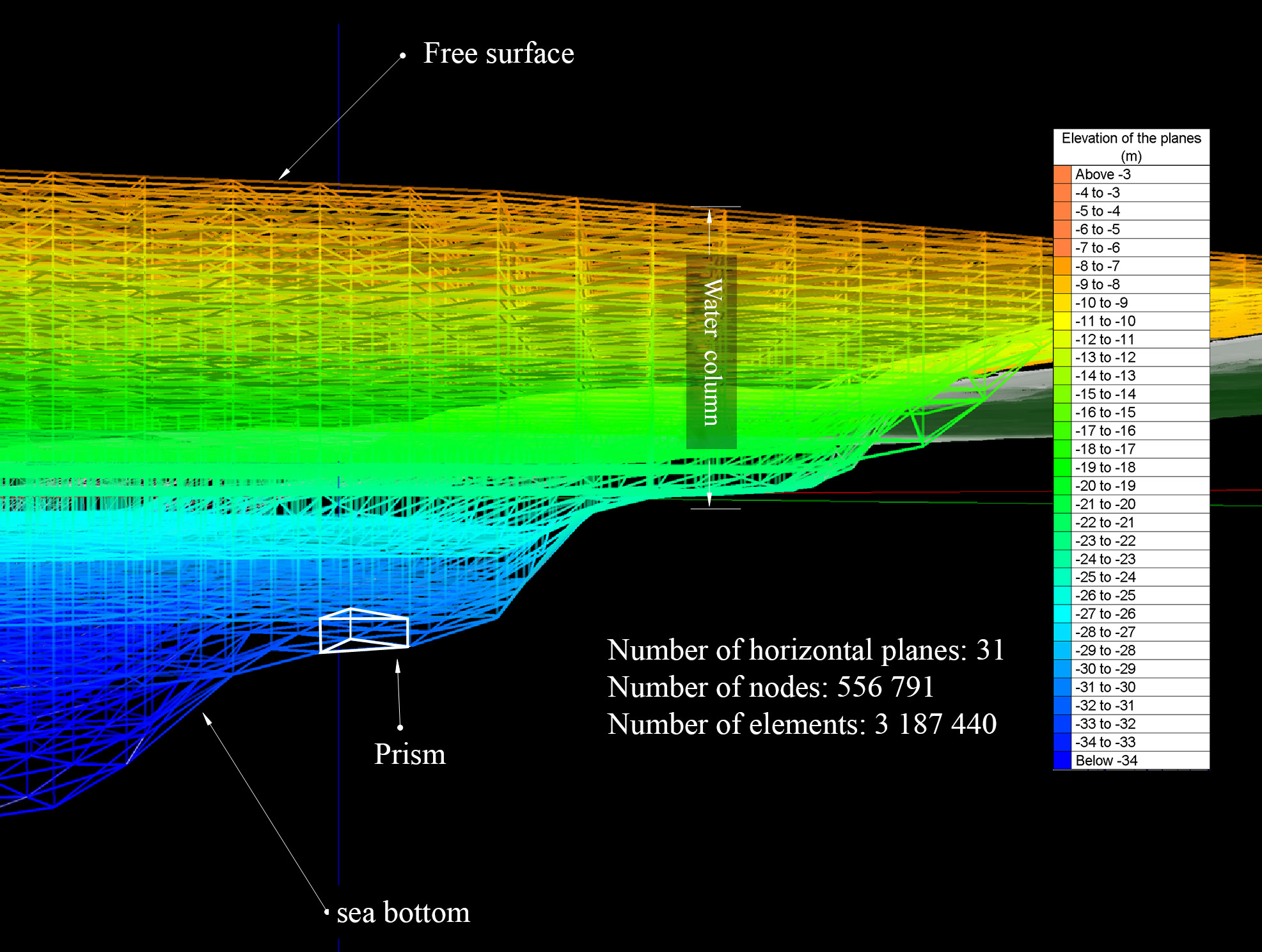}
	\caption{Computational 3D grid of Nice coastal area.}
	\label{Mesh3D}
\end{figure}

The system of equations (1-4) is solved with the Finite Element Method applied to a spatial discretization using prisms. Each prism has six nodes and a vertical quadrangular side \cite{Hervouet07}. The grid domain for Nice has been constructed with the Blue Kenue mesh generator to create a 2D horizontal mesh \cite{NRC10}. Two zones of different mesh size were constructed on this triangular unstructured grid. The first zone is defined by a resolution of 3m and is centered on the sewage outlet. It includes all the observation points. The second zone is made of 8m triangals and covers the remaining part of the study domain (fig. \ref{Mesh2D}). This 2D mesh has been repeated over the vertical 30 times in order to build the three-dimensional mesh composed of 3187440 elements and 556791 nodes. The vertical density of the mesh is higher around  the 40 first meters depth where the wastewater are discharged (Fig. \ref{Mesh3D} and \ref{Sampling-points}).\\

The three-dimensional hydrodynamic model is coupled with the wave propagation code TOMAWAC (TELEMAC-based Operational Model Addressing Wave Action Computation) \cite{Benoit96}.
This phase averaged model calculates the sea states by solving the balance equation of the action density directional spectrum  at each point of the horizontal computational grid described previously \cite{Goda00},\cite{Komen94}. The wave driving forces computed in TOMAWAC are transmitted to TELEMAC3D which calculates the nearshore current \cite{Tomawac11}. This free and opensource integrated suite of solvers have proved its accuracy on several nearshore applications \cite{Aelbrecht99}. 

In order to set the initial wave and current conditions, a hotstart file containing the modelling sea state observed just before the beginning of the experiment is used. The study domain is initially free of any discharge coming from the sewage network. So the initial driving forces are only the wave field and the wave induced current.

On the boundary, a parameterized Jonswap spectrum \cite{Hasselmann73} is set up based on the measurements from Meteo-France wave buoy located 60km from the coast and transferred to the domain limits. The velocities on the outline of the studied area are supposed nil because no current observations was available on a larger scale. 
To compute heat exchange between water and the atmosphere, a linearised formula expressing the balance of heat exchange fluxes at the free surface is chosen \cite{Sweers76}
\begin{equation}
K\frac{\partial T}{\partial z}=\frac{A}{\rho C_{p}}(T-T_{air})
\label{heatEx}
\end{equation}
where $K$ is the coefficient of moleculare heat diffusion in water, $C_{p}$ is the water specific heat equal to 4140 $J/kg/K$.
$A$ is the exchange coefficient in $W/m^{2}/K$ and is expressed according to the temperature of water $T$ and the wind velocity $u_{wind}$ in $m/s$:
\begin{equation}
A=(4.48+0.049T)+2021.5 b(1+u_{wind})(1.12+0.018T+0.00158T^{2})
\label{paramA}
\end{equation}
where $b$ is a parameter depending on the location of the study. In the Mediterranean it can be set to 0.0035.

Concerning the numerical scheme, the implicit Finite Element Method is solving the diffusion step of tracers and velocities. Regarding the advection of velocities the characteristics method is used and for the advection of tracers the explicit Finite Element Method is applied.

\subsection{\textbf{The measurement campaigns methodology}}
Two measurements campaigns were designed in order to quantify the reliability of the simulation built using the TELEMAC system. Both of them are based on the same pollution events specific to Nice.
Each year, in November, the main pipe of the sewer system of Nice is completly cleaned up and checked up by the municipality. These operations are carried out by zones and points of interest. In order to protect the workers, the sewages collected by the section under maintenance are deviated through the marine outfalls. Thus, the pollution event is completly planned and controlled (each outfall is monitored with discharge measurement stations).
As a consequence, these events are perfectly adapted to validate our modelling work.

Based on logistical limitations of Nice Municipality, it has been decided to concentrate our efforts on only one outfall: la Païole. Two main criteria have oriented our choice. The first one was the depth of the outfall (30m). At this depth, the effect of the density gradient on the transport of the polluant can be observed. The second criterion is based on the isolation of the outfalls (130m from the coast). In the area of La Païole, there are no other pollution sources, so only one outfall can be monitored.

\begin{figure}
	\centering
		\includegraphics[width=14cm]{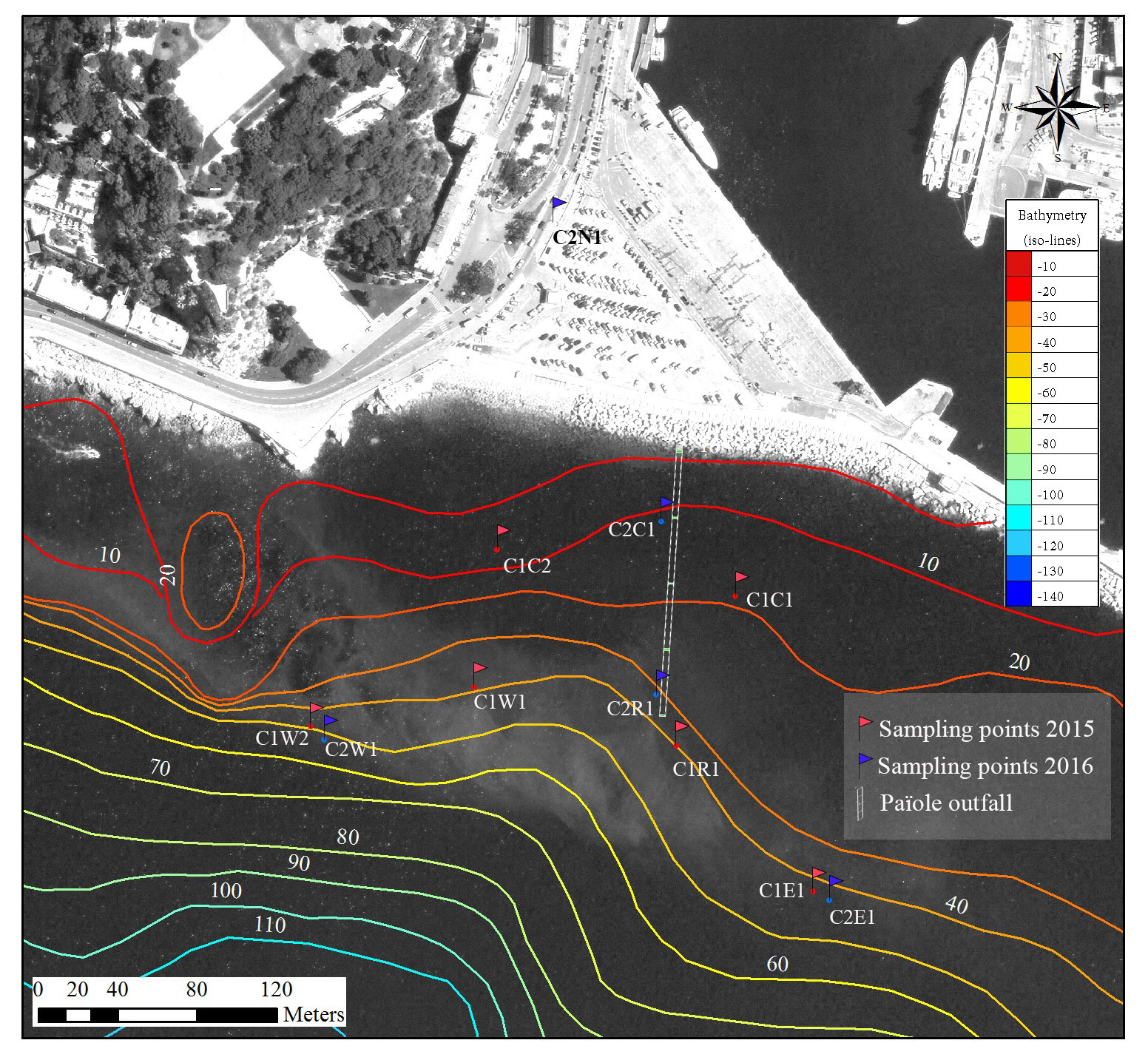}
	\caption{Sampling points location-2015 and 2016 campaign.}
	\label{Sampling-points}
\end{figure}

Four sampling points were defined to follow the spatial distribution of the bacteria concentration: one right at the outlet, one further north near the coast, and two to the east and west (fig. \ref{Sampling-points}). 

\begin{figure}[!h]
	\centering
		\includegraphics[width=10cm]{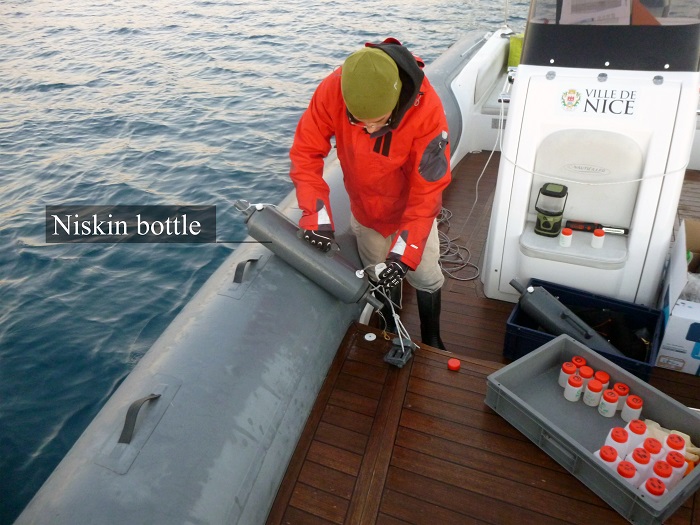}
	\caption{Sampling  using a Niskin bottle.}
	\label{sampling}
\end{figure}

Regarding the 2015 experiment, at each sampling point, three measurements were taken along the water column using Niskin bottles (fig. \ref{sampling}): at the surface, in the middle and at the bottom of the water column (around -1, -10 and -25 meters depth). These samples were taken at 9:00 PM, midnight, 3:00 AM and 8:00 AM to follow the time evolution of the pollutant concentrations. 

The 2015 campaign has raised some issues related to this protocol. Firstly, during the campaign, we observed that the outfall was not at the expected position. So we moved the 4 sampling points initialy planned during the night (fig. \ref{Sampling-points}). Secondly the data of flows discharged in the sea were known, but not the bacteria concentration of the raw sewage. And finally the analysis of the samples shows that the pollution propagated only at the surface. This observation will be discussed and explained in the following part.

As a consequence, in November 2016 a new campaign was conducted with an improved protocol. The sampling points were revised based on the new location of the outlet on the map and samples were only collected at the surface. The sampling schedule was also modified for a better capture of the extrema ( 9:00 PM , midnight, 2:00 AM and 6:00 AM). In order to obtain the bacteria concentration of the raw sewage, samples were taken directly from the network upstream of la Païole at  8:00 PM  and 3:00 AM.

In 2015 and 2016, the samples were analyzed by the municipality using a method derived from the most probable number method. The latter consists in inoculating numerous dilutions of the sample in enrichment broth tubes. Subsequently, the tubes are transplanted onto selective agar allowing, depending on the number of positive tubes and their distribution, to estimate the most probable number of bacteria (MPN) \cite{McGrady18}, \cite{Halvorson32}.
Thus, the sample is diluted and then seeded in a series of wells of a microplate which contains the dehydrated culture medium. The microplates are examined under UV after 36 to 72 hours of incubation at the temperature $317.15K$. The presence of Escherichia coli and Enterococci is revealed by detection of their enzyme \cite{COFRAC08}.

\newpage
\section{Results}
\subsection{\textbf{Experimental results}}
\begin{figure}[!h]
	\centering
		\includegraphics[width=12cm]{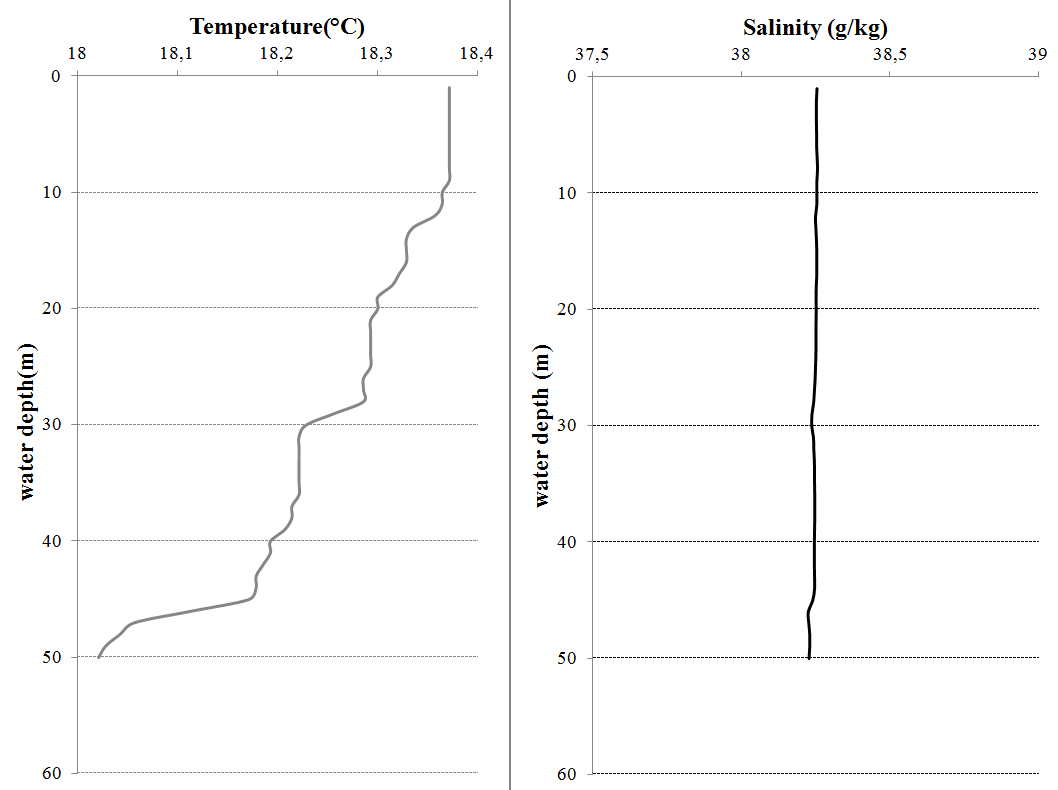}
	\caption{Water temperature and salinity measured}
	\label{Temp_salinity-2015}
\end{figure}

In order to set up the 3D model, the temprature and salinity were followed during the experiment thanks to the Eol Buoy located in Villefranche bay, 6km East of Nice.
This sampling tool is able to measure the two physical parameters from the depth of 100m to the surface every hour.
The experiments took place in autumn during which the climate is characterised by an air temprature around $293.15K$, 10 degrees less than in summer. 
So the temperature profile is evolving from a stratified state to an homogenous state with values of $291.15K$ at 50m depth and $291.5K$ at the surface (fig. \ref{Temp_salinity-2015}).
Regarding the salinity, the variations observed are not significant. So the salinity profile can be considered as homogeneous on the studied domain (fig. \ref{Temp_salinity-2015}). \\

The outlet discharges were carried out using a Détronic automatic captor. The discharge event started at 7:00 PM and finished around  4:30 AM . During 8 hours,  5300 $\text{m}^3$ of raw sewage were released in the sea for both events. During the night the hourly mean is decreasing from 220 l/s to 145 l/s. This evolution illustrates the changes of the individual wastewater-generating activities of this residential dwelling \cite{USEPA}. The variability of the appliance usage is also characterised by instantaneous peak flows reaching 260 l/s.

\begin{figure}[!h]
	\centering
		\includegraphics[width=12cm]{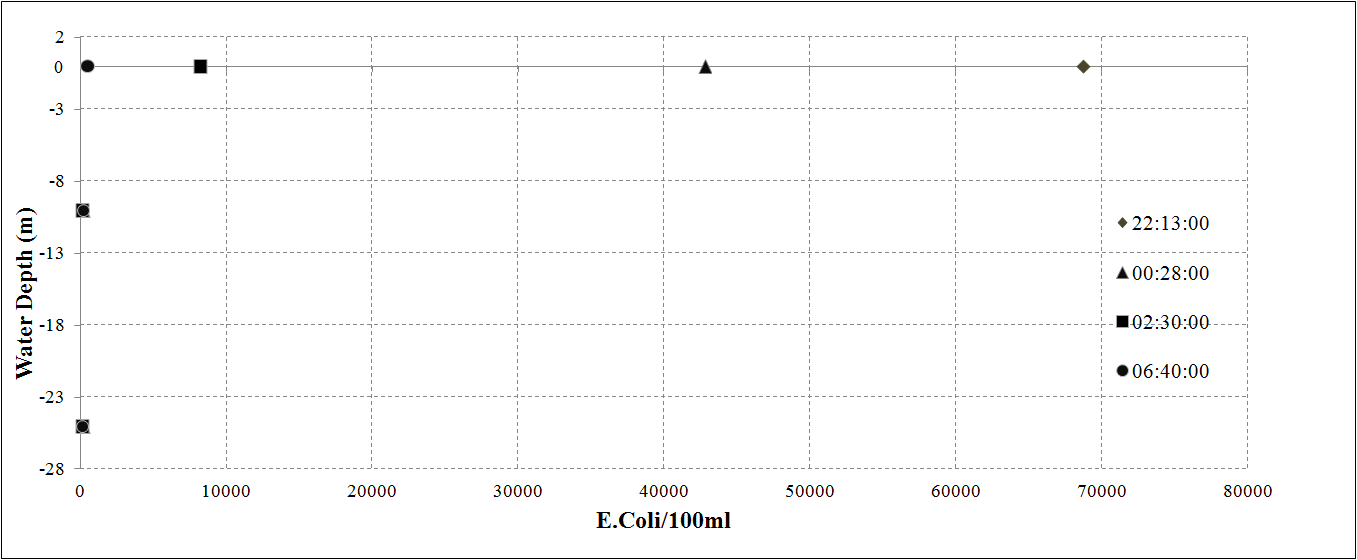}
	\caption{\textit{Escherichia coli} concentration measured through the water column on the C1R1 sampling point}
	\label{EcoliDepthProfile}
\end{figure}

The E.coli concentration results of the 2015 campaign have demonstrated that the bacteriological impact of the raw sewage discharge is spreading through the surface without any contamination of the deeper water (fig. \ref{EcoliDepthProfile}). This phenomenon comes from the density of the marine environment and the effluent. In fact, the density of the raw effluent is lower than the salt water described previously because of its temperature and salinity. \\

\begin{figure}[!h]
	\centering
		\includegraphics[width=12cm]{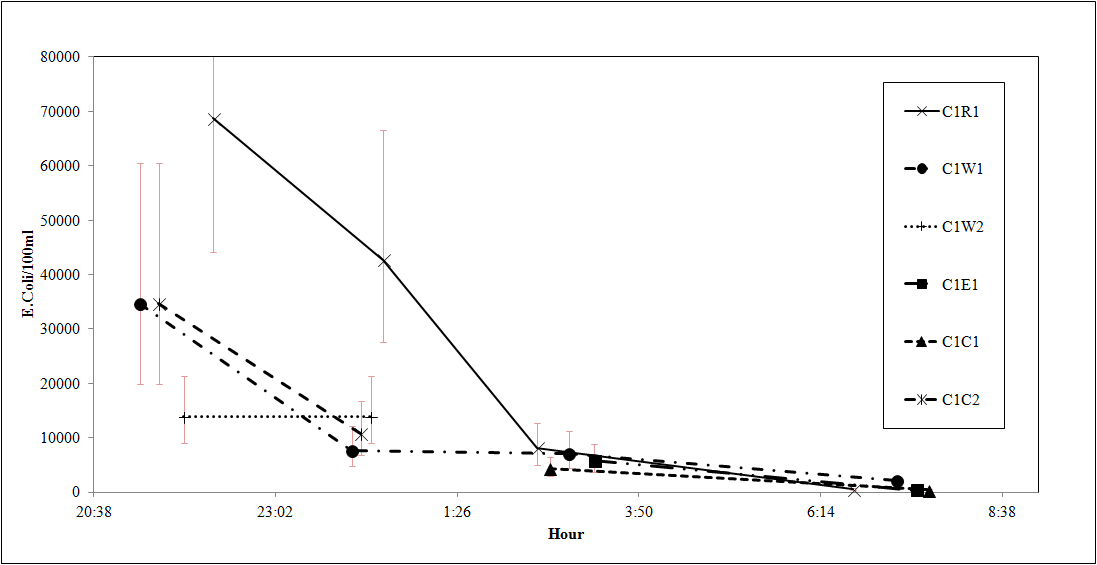}
	\caption{\textit{Escherichia coli} concentration measured on each sampling points of the 2015 sampling campaign at the free surface - the vertical error bars represent the 95\% confidence interval of the Most Probable Method}
	\label{ExperiResult15}
\end{figure}

Regarding the surface results (Fig. \ref{ExperiResult15}), the C1R1 gives the highest bacteriological counts due to its proximity with the outlet. The first concentration measured reaches 70 000 E.Coli/100ml which is 70 times higher than the legal threshold. During the night, the bacteriological count is decreasing until 10 000 E.Coli/100ml at 2:00 AM and 150 E.Coli/100ml at 6:10 AM. The end of the flow, the mortality and the diffusion/advection phenomenon explain the concentration measured during the morning. Nevertheless, the decrease observed between  7:00 PM  and 2:00 AM is not correlated with the discharge values which stay high during the event.

Around the outlet, the sewage plume is spreading in all directions. Even if the concentrations are lowered, the west, the east and the nearshore points are following the same patterns than the C1R1 sampling point.  

\begin{figure}[!h]
	\centering
		\includegraphics[width=12cm]{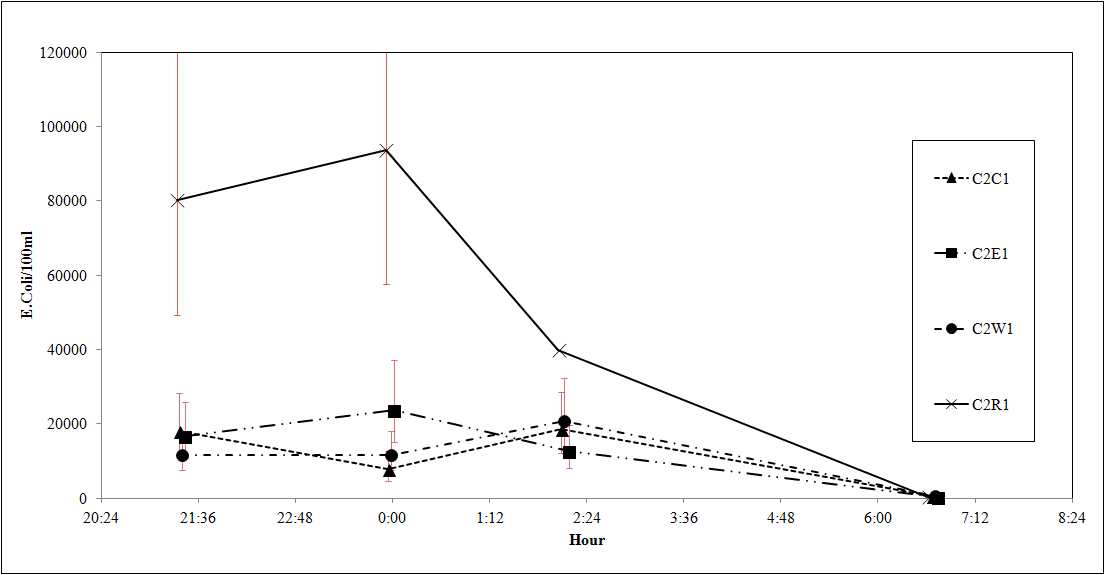}
	\caption{\textit{Escherichia coli} concentration measured on each sampling points of the 2016 sampling campaign at the free surface - the vertical error bars represent the 95\% confidence interval of the Most Probable Method}
	\label{ExperiResult16}
\end{figure}

In 2016, the order of magnitude of the bacteriological counts were equal to the 2015 campaign and the behaviour of the pollutant plume distribution was similar to the previous description (fig. \ref{ExperiResult16}).
 
\begin{figure}[!h]
	\centering
		\includegraphics[width=7cm]{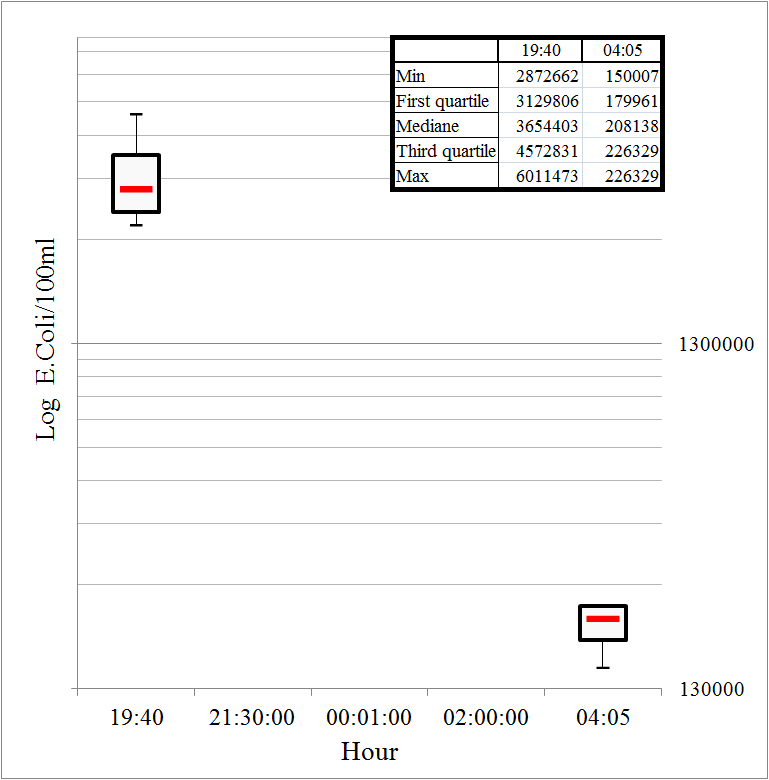}
	\caption{\textit{Escherichia coli} concentration measured on the network sampling point, 2016 campaign - box plot based on four replicates for each sample}
	\label{ExperiResulNet16}
\end{figure}

The raw sewage samplings (C2N1) results demonstrate the variability of the wastewater quality during the night (fig. \ref{ExperiResulNet16}). The important decrease of the feacal contamination are due to the individual water-uses events \cite{UnivWis78}. In fact, the water uses pattern changes drasticly from  7:00 PM  to 3:00 PM \cite{Anderson89}, \cite{Anderson93}, \cite{Brown84}, \cite{Mayer99}.
These changes in the wastewater quality are at the origin of the decrease of the bacteriological counts observed mainly on the sampling points C1R1 and C2R1.

\newpage
\subsection{\textbf{Modelling results}}
The 3d model results are compared to the experiments results. The E.Coli concentrations at the outlet of the model were set using the wastewater quality results discussed previously.

Firstly, the vertical diffusion and advection phenomena are well represented in the simulations. In fact, the model calculates the right bacteria concentration on the surface with a vertical propagation limited to 5 meters. This local effect explains why no trace of bacteria were found on the sampling point C1W1, C1W2, C1E1 and C1R1 at 10m and 25m depth (fig. \ref{EcoliDepthProfile}). 

\begin{figure}[!h]
	\centering
		\includegraphics[width=10cm]{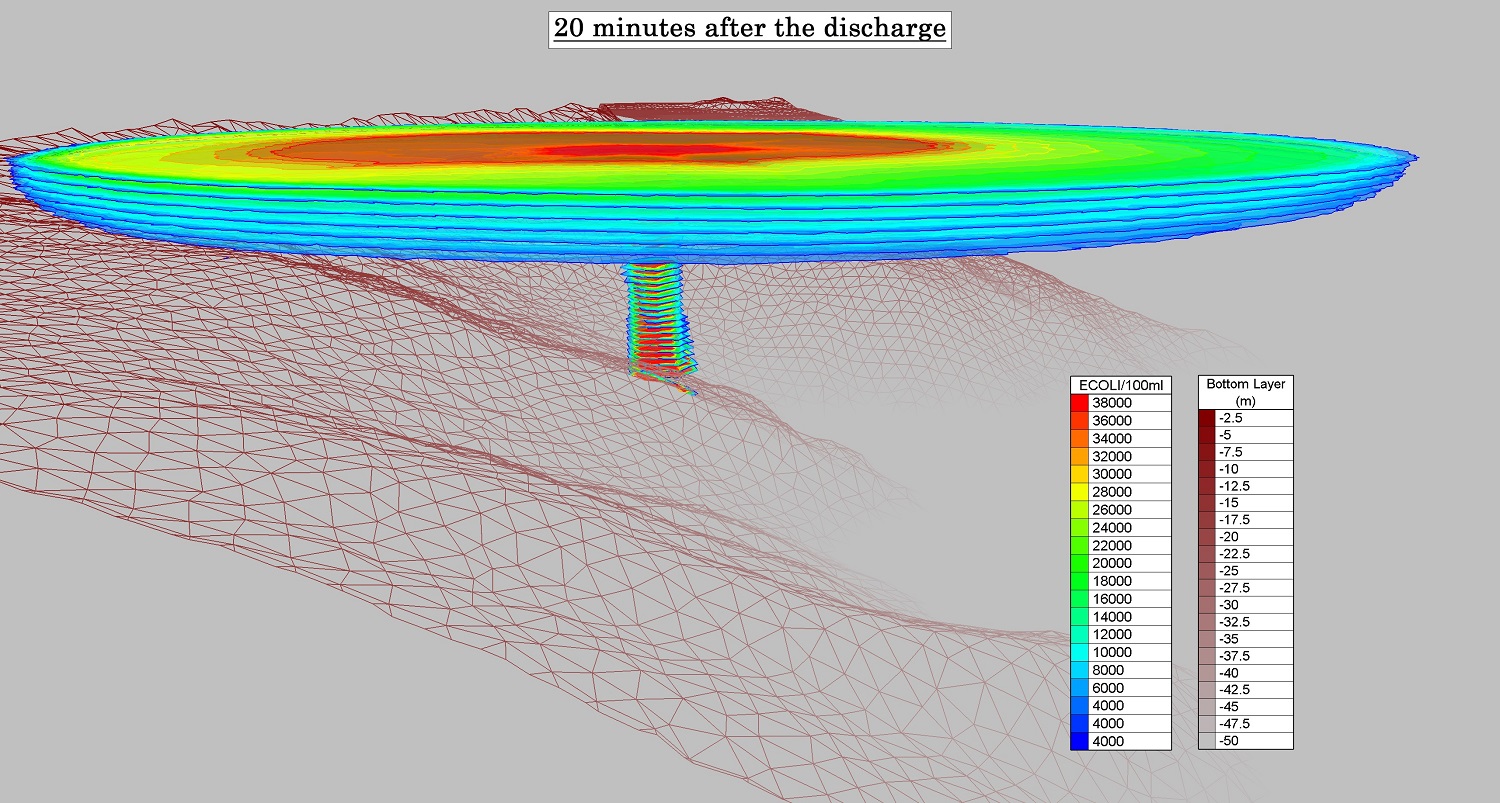}
	\caption{3D field of \textit{Escherichia coli} concentration around the sewage outlet}
	\label{ModResulEcol16}
\end{figure}

\begin{figure}[!h]
	\centering
		\includegraphics[width=10cm]{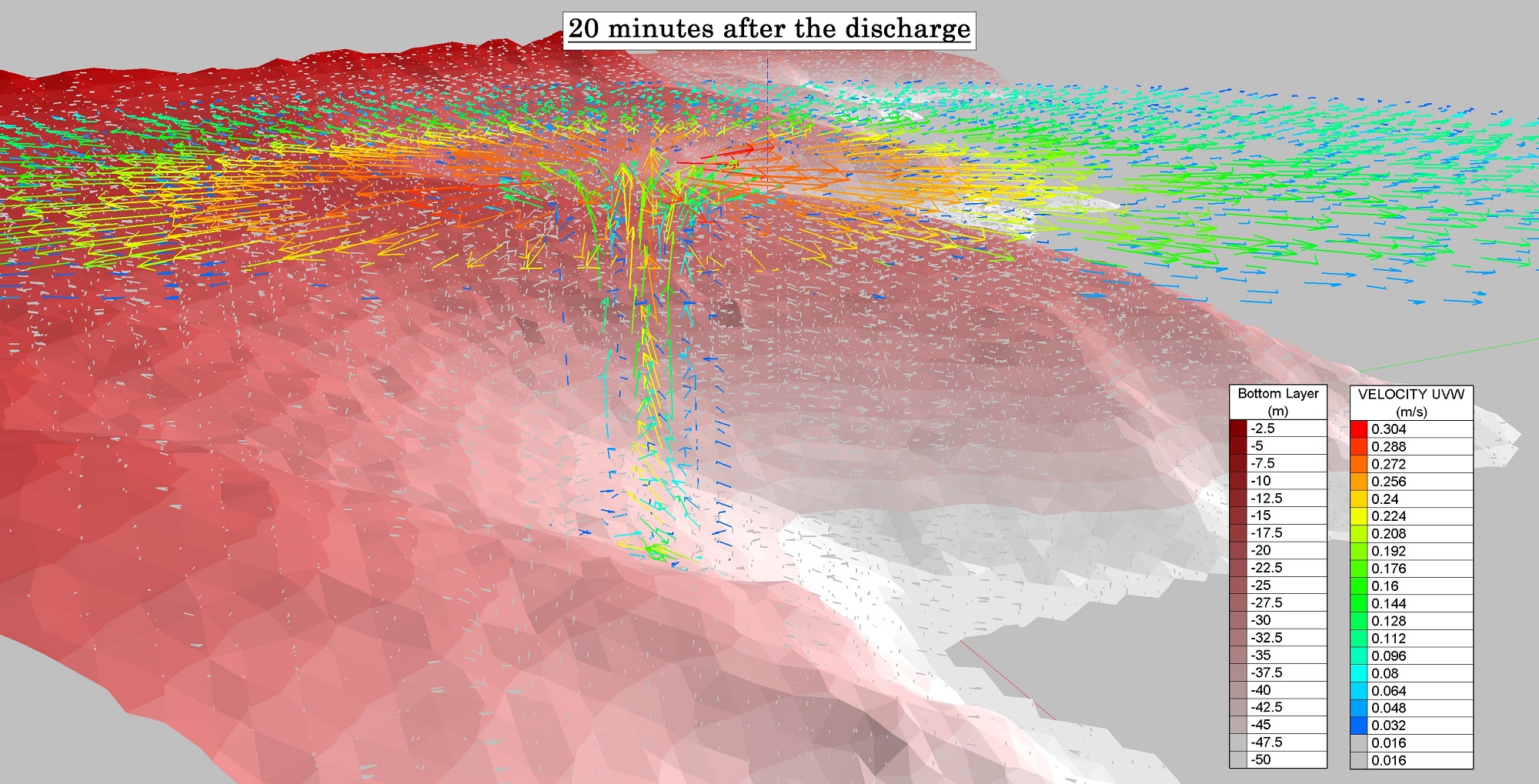}
	\caption{3D velocity vector field around the sewage outlet}
	\label{ModResulVit16}
\end{figure}

\begin{figure}[!h]

	\centering
		\includegraphics[width=14cm]{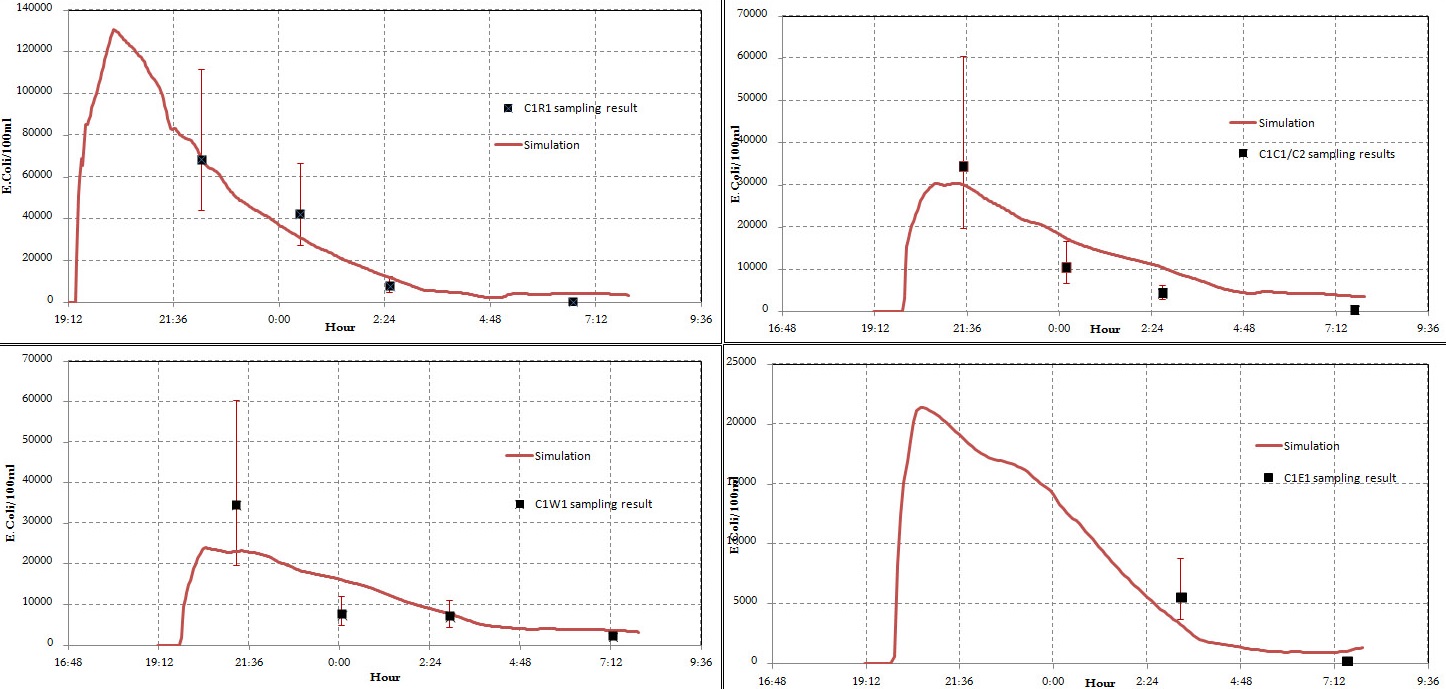}
	\caption{\textit{Escherichia coli} concentration simulated and observed at the free surface, 2015 campaign - the vertical error bars represent the 95\% confidence interval of the Most Probable Method}
	\label{ModelResult15}
\end{figure}

Moreover, the 2015 E.Coli concentration simulated tends to fit the experiments measurements for the East, West, North points which means that the horizontal diffusion and advection terms are well described (fig. \ref{ModelResult15}).

\begin{figure}[!h]
	\centering
		\includegraphics[width=12cm]{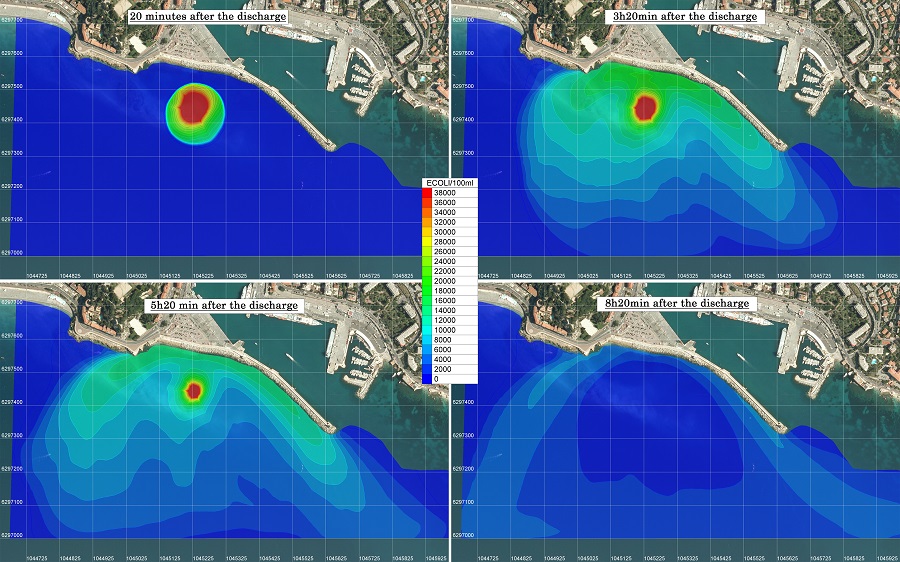}
	\caption{Time dependent 2D fields of \textit{Escherichia coli} concentration at the free surface}
	\label{EcoliSurface}
\end{figure}

\begin{figure}[!h]
	\centering
		\includegraphics[width=12cm]{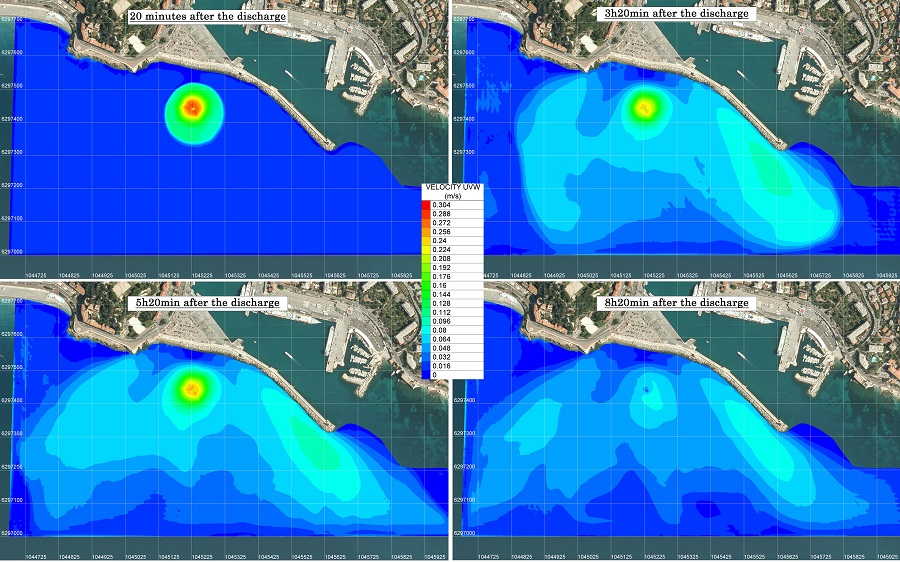}
	\caption{Time dependent 2D fields of velocity at the free surface}
	\label{VitSurface}
\end{figure}

\begin{figure}[!h]

	\centering
		\includegraphics[width=14cm]{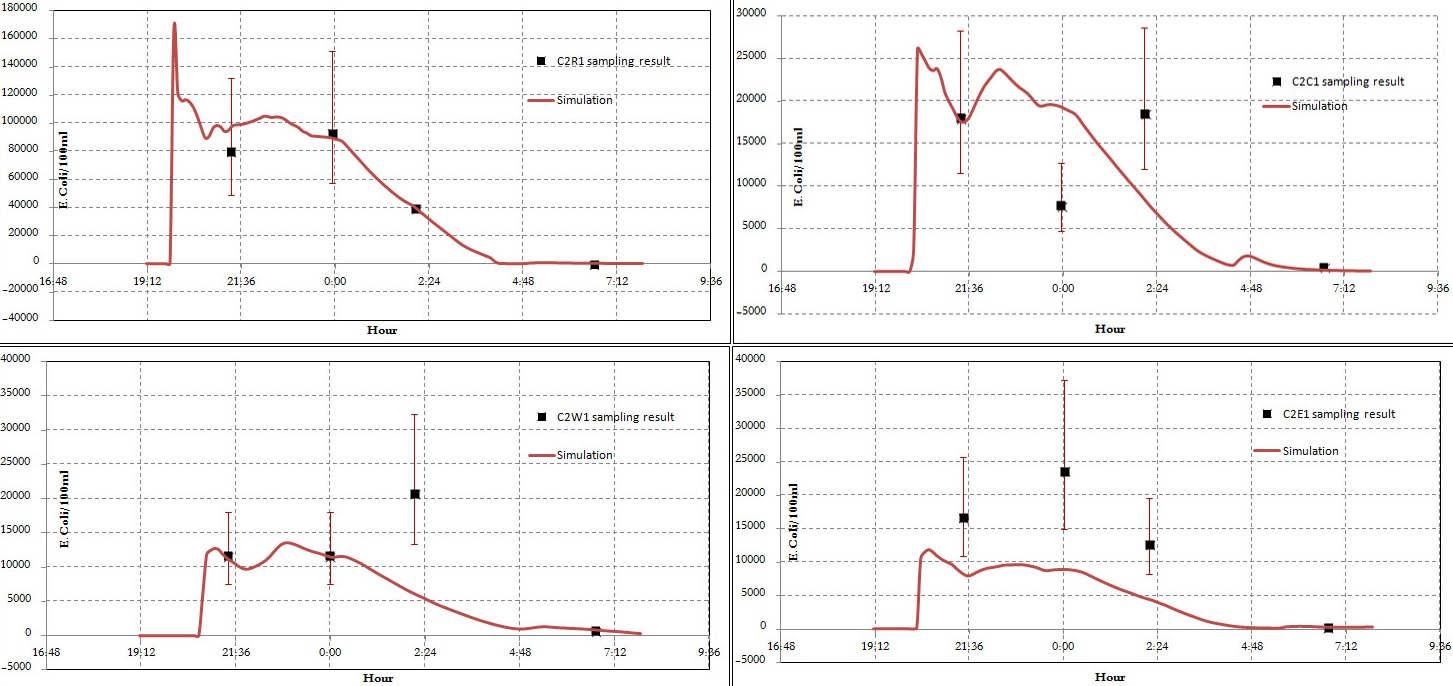}
	\caption{\textit{Escherichia coli} concentration simulated and observed at the free surface,  2016 campaign - the vertical error bars represent the 95\% confidence interval of the Most Probable Method}
	\label{ModelResult16}
\end{figure}

Regarding the 2016 sampling campaign, the simulation shows poor results on the East point C2E1 (fig. \ref{ModelResult16}). The underestimation of the E.Coli counts could come from the weakness of this local scale to represent the effect of the wind or currents on the transport of the pollutant. But the lack of current and wave data obervations makes a deeper validation of the hydrodynamic driving forces more complicated.
  
\noindent Whereas the model satisfactorily replicates the evolution trend of the other sampling points, it fails to capture the increase of E.Coli concentration at around 2:00 AM observed on C2C1 and C2W1 (fig. \ref{ModelResult16}). The C2R1 experimental results show a constant decrease of the bacterial concentrations from  9:00 PM  to 7:00 PM, emphasizing the inconsistency of the 2:00 AM increase on C2C1 and C2W1.  

\noindent Consequently, further analysis of the C2C1 samples were carried out by counting the intestinal enterococci. The results of this microbiological indicator inidicates a constant decrease of the concentrations during the night following the C2R1 pattern.

\subsection{\textbf{Mesh sensitivity}}

Three configurations have been tested to study the mesh resolution sensibility. In order to optimize the calculation duration, unstructured grids with variable mesh sizes have been chosen. The first configuration is built with 10m elements in a circle of 120 m around the sewage outlet. Then, the mesh size is increasing gradually to reach 30m at the observation points C2W1 and C2E1.
The same setting has been implemented for the second configuration with a minimum grid size around the outlet of 5 meters. And finally the last configuration is the one introduced in section 2.1, the grid size is set to 3m, uniformly from the pipe to the observation points.

\begin{table}[!h]
	\centering
		\begin{tabular}{|l|c|r|}
			\hline
			Low resolution grid & Medium resolution grid & High resolution grid \\
			\hline
			Grid size between 10m and 30m & Grid size between 5m and 30m & Grid size 3m \\
			\hline
			\includegraphics[width=4cm]{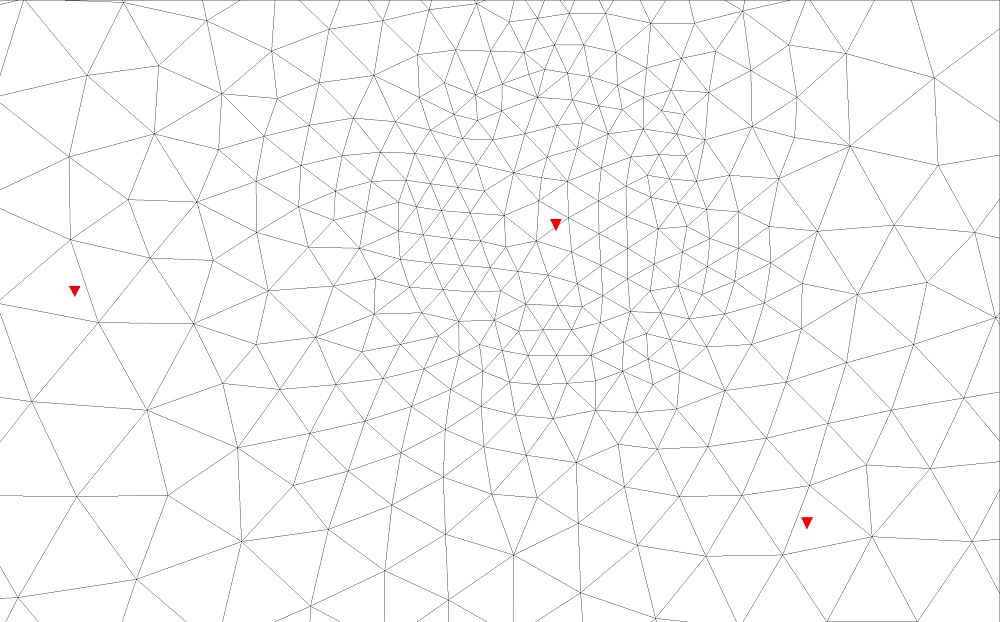} & \includegraphics[width=4cm]{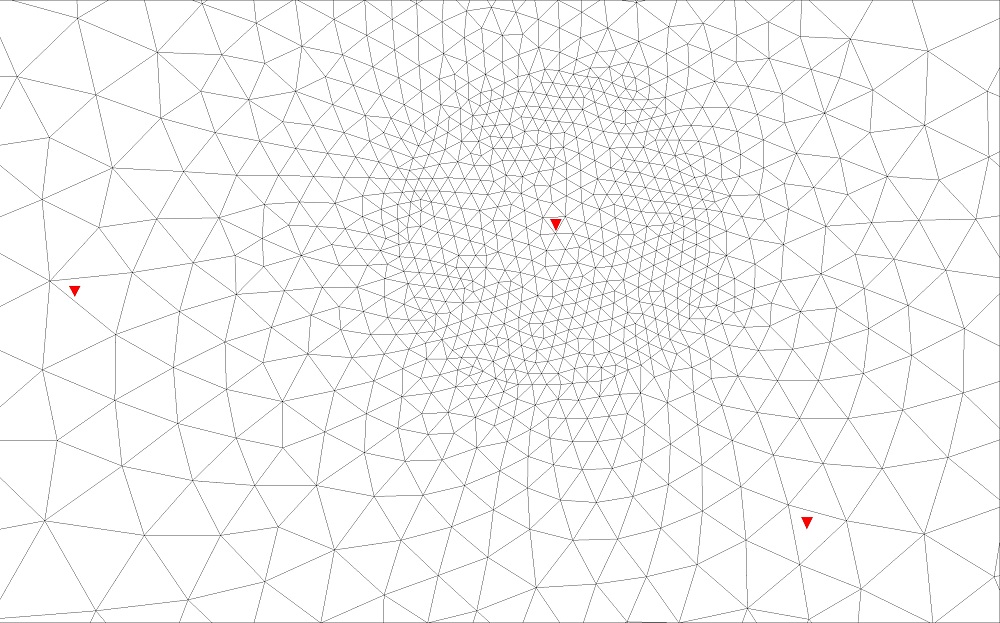} & \includegraphics[width=4cm]{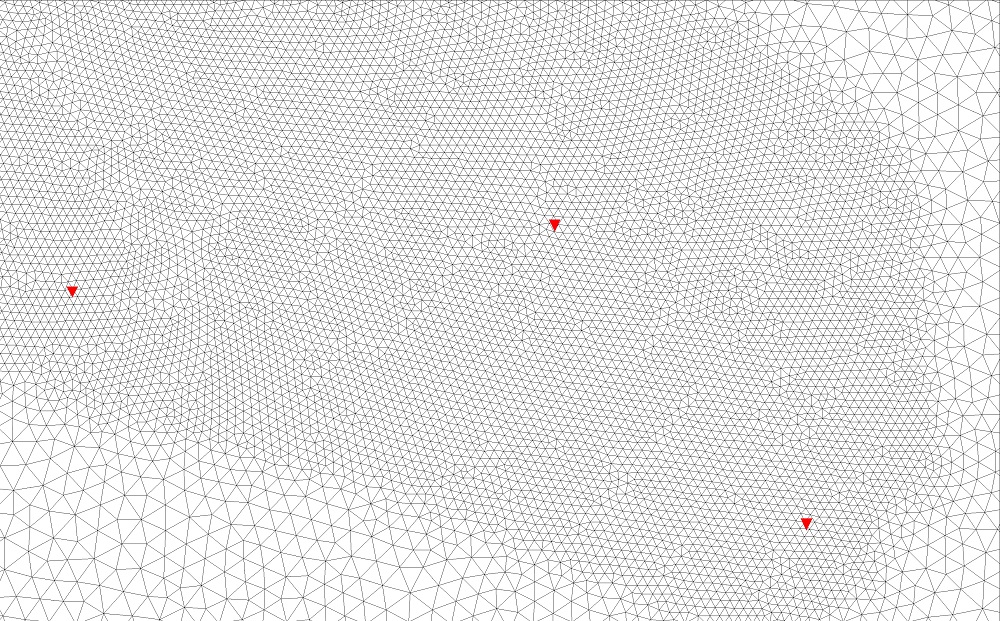} \\
			\hline
		\end{tabular}
	\caption{Description of the grids tested for the sensitivity analysis - in red triangles the 2016 observation points}
	\label{Tabgrid}
\end{table}

\begin{figure}[!h]
	\centering
		\includegraphics[width=10cm]{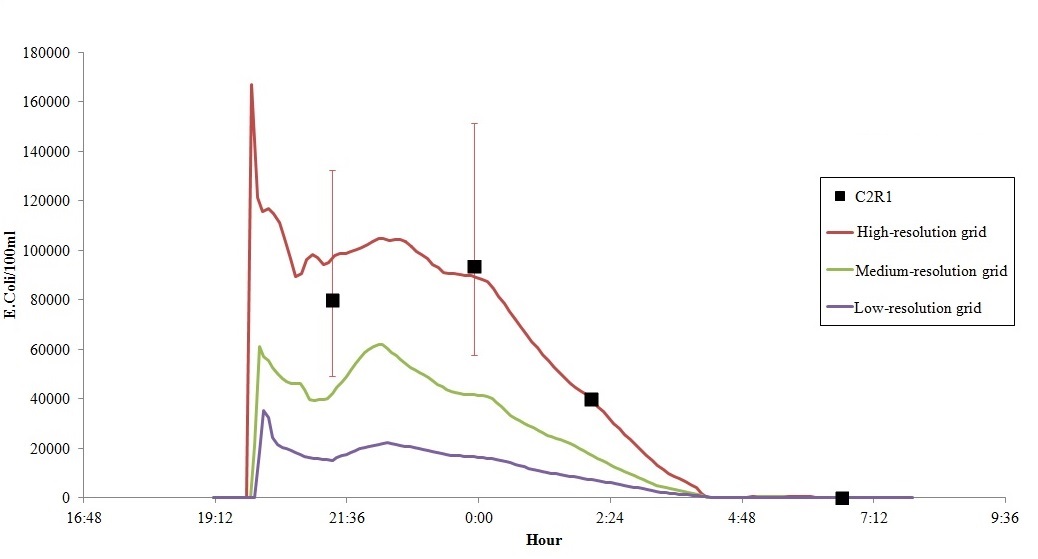}
	\caption{\textit{Escherichia coli} concentration simulated and observed at the free surface on the point C2R1-all grids, 2016 campaign}
	\label{MeshSensib}
\end{figure}

\begin{figure}[!h]
	\centering
		\includegraphics[width=10cm]{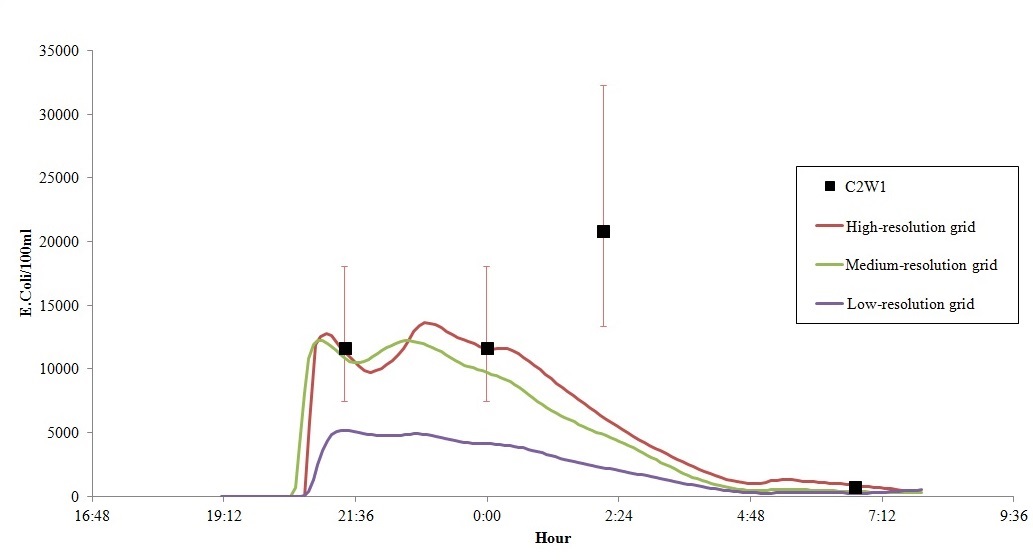}
	\caption{\textit{Escherichia coli} concentration simulated and observed at the free surface on the point C2W1-all grids, 2016 campaign}
	\label{MeshSensib2}
\end{figure}

\begin{figure}[!h]
	\centering
		\includegraphics[width=10cm]{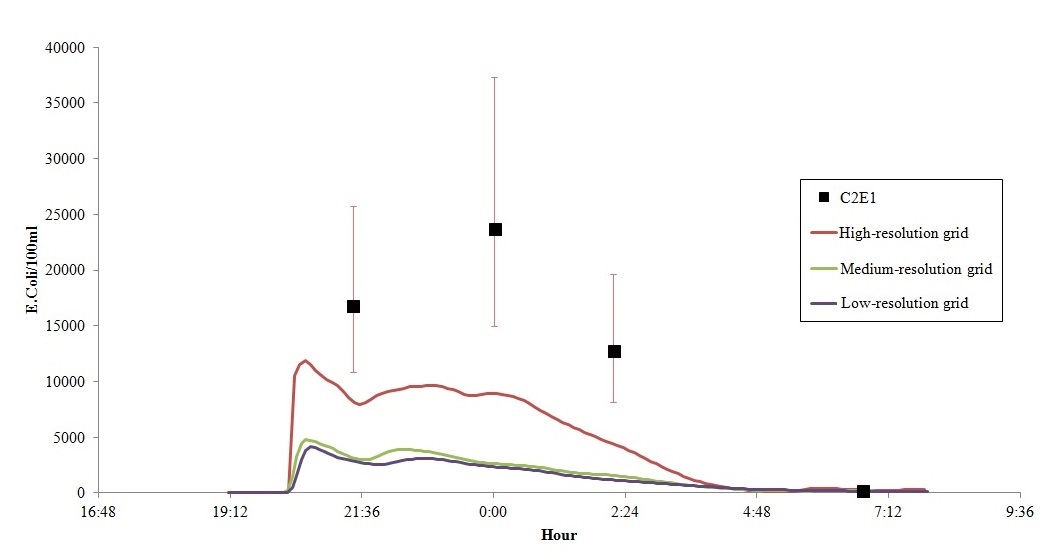}
	\caption{\textit{Escherichia coli} concentration simulated and observed at the free surface on the point C2E1-all grids, 2016 campaign}
	\label{MeshSensib3}
\end{figure}

Figures \ref{MeshSensib},\ref{MeshSensib2} and \ref{MeshSensib3}  show that the accuracy of the simulation is increasing with the grid resolution. This result can be achieved because the numerical diffusion is reducing gradually with the mesh size.  
Nevertheless the C2E1 observation concentrations are not reached even with the high resolution grid. Further simulations should have been tested with higher resolution in order to improve these results and to reach the mesh convergence but the calculation capacity of Nice Municipality is limited. 
In fact, the final goal of this project will be to create an operational tool that will be able to simulate a bacteriological pollution to inform decision maker during pollution events. Thus, the model should give accurate results quicker than the bacteriological laboratory analysis (36h). A number of elements more important than the one in the 3m resolution grid would give a computation time higher than 36hours based on the computational infrastructure of Nice municipality.


\newpage

\section{Discussion}
Two major points were stressed out by the experiment:

\begin{itemize}
	\item The uncertainty of the counting method for the microbiological indicators is high \cite{Man75}. In fact, the samples were treated based on the most probable number of bacteria in the volume of sea water. For each sample, the confidence limits are illustrated on the graphs by red segments (fig. \ref{ModelResult15}, \ref{ModelResult16}). As observed, the higher the concentration of bacteria is, the more the uncertainity increases. Consequently, the extremly high values observed in the raw effluent samples have to be associated with a large range of possible values explaining the quantiles, the minimum and the maximum illustrated on (fig. \ref{ExperiResulNet16}). This observation impacts directly the calibration and validation process of a bacterial quality model. The input values of concentration that have to be imposed on the source points (outlet, rivers) are mostly unknown or associated with high uncertainties.  

	\item Moreover, the variability of the pollution in the network during the night has been demonstrated in this study. Only two hours (21h, 3h) were investigated. But we can assume that the bacterial count in the network is fluctuating every hour due to the changes in the uses. This hourly variability can explain the differences observed on the model results in which the decrease of the network bacteriological count is supposed to be constant.\\ In order to accuratly predict the evolution of the bacterial pollution, a deep study of the raw effluent is needed. Several temporal scales must be investigated: hourly and dayly in order to characterise the variation of the E.Coli concentration due to the individual wastewater-generating activities, and seasonaly or based  on meteorological events to evaluate the extreme values.

\end{itemize}

This paper aims at comparing 3D hydrodynamic simulations of coliforms tranport and decay to in situ observations. Compared to other similar studies \cite{Gupta05},\cite{Huang95}, \cite{Liu07}, \cite{Thupaki10}, the choice of these indicators, the frequency of the measurments and the spatial distribution of the monitored points allow to describe the pollution evolution and its transport at a local scale.  Even if these microbiological parameters are the main focus of the public decision makers, the performance of the model should also be tested with physical and chemical data such as the salinity, the temperature, the current velocities and the wave heights. Because of the lack of measurement instruments, this work has not been achieved yet.

\section{Conclusion}
In a marine environment, the bacteriological pollutions depend on chemical and physical parameters which interact all along water column (salinity, temperature, light intensity, wave actions). As a consequence , it was necessary to set up specific measurement campaigns in order to calibrate and validate a 3D model based on OPENTELEMAC\\
In 2015 and 2016, two campaigns have allowed us to get a better understanding of the bacterial pollution behavior in this field. Their protocol were funded on spatialy distributed measures of microbiological indicators around the outlet and upstream in the network.  Added to that, the water column chemical and physical parameters were gathered thanks to a measurement buoy located 50 km away from the pollution source. These experiment results are an ideal data set to develop a three dimensional bacterial quality model because the vertical and horizontal diffusion/convection phenomenon around the source point are clearly identified.  They also permit to quantify the variability of the network raw effluents. Even if the data collected must be completed with local physical and chemical observations, the strength of this approach is its reproducibility. Indeed, the same pollution event occurs every year so a more complete campaign can be designed.\\
The Reynolds-Averaged Navier-Stokes equation associated to the non-hydrostatic pressure assumption successfully reproduce the bacterial concentration evolution in the marine environment. Nonetheless, the high confident limits linked to the laboratory analysis of the E.Coli concentrations make the calibration and validation processes complex. Based on these uncertainities, the TELEMAC3D model developped in this paper shows satisfactory results. And more importantly, the tool has sucessfully determined the end of the pollution for both events. This last aspect is a main point in the process of creating a prediction tool for management crisis purpose. 

\section{Acknowledgement}
The authors declare that they have no conflict of interest.

We thank Doctor Luisa Passeron-Mangialajo from Laboratory Ecology and Conservation Science for Sustainable Seas of the University of Nice for her assistance with sampling methodology using Niskin bottle.
We would also like to show our gratitude to Doctor Grisoni Jean Michel from the Institute of the Sea of Villefranche, CNRS, for sharing the data of the buoy Eol developped by its research team. 
We thank Julien Larraun from Metropole Nice Côte d'Azur for his help on the simulation work.





\end{document}